# Dependability Tests Selection Based on the Concept of Layered Networks

Andrey A. Shchurov, Radek Mařík

**Abstract**— Nowadays, the consequences of failure and downtime of distributed systems have become more and more severe. As an obvious solution, these systems incorporate protection mechanisms to tolerate faults that could cause systems failures and system dependability must be validated to ensure that protection mechanisms have been implemented correctly and the system will provide the desired level of reliable service. This paper presents a systematic approach for identifying (1) characteristic sets of critical system elements for dependability testing (single points of failure and recovery groups) based on the concept of layered networks; and (2) the most important combinations of components from each recovery group based on a combinatorial technique. Based on these combinations, we determine a set of test templates to be performed to demonstrate system dependability.

**Index Terms**— Dependebility testing, distributed systems, formal models, layered networks.

————————— ◆ —————————

## 1 INTRODUCTION

COMPUTING systems have come a long way from a single processor to multiple distributed processors, from individual-separated systems to networked-integrated systems, and from small-scale programs to sharing of large-scale resources. Moreover, nowadays virtualization and cloud technologies make another level of distributed system complexity. On the other hand, the consequences of failure and downtime have become more severe. They might endanger human lives and the environment, do serious damage to major economic infrastructure, endanger personal privacy, undermine the viability of whole business sectors and facilitate crime [1]. As an obvious solution, computing systems incorporate protection mechanisms to tolerate faults that could cause systems failures and, as a consequence, the most difficult part of systems deployment is the question of assurance that system dependability mechanisms (fault tolerance or high availability) have been implemented correctly and a system is able to provide the desired level of reliable service.

Dependability is an integrating concept that unites the attributes of reliability, availability, safety, integrity and maintainability. The original definition of dependability determines the system ability to deliver service that can justifiably be trusted [2] (this definition stresses the need for justification of trust). The engineering definition is simpler – dependability is the ability of a system to avoid service failures or the probability that a system will operate when needed. The four major categories of dependability are:

- fault prevention, i.e. prevention of the occurrence or introduction of faults;
- fault removal, i.e. reduction of the number and severity of faults;
- fault forecasting, i.e. estimation of the present number and the likely consequences of faults
- fault tolerance, i.e. avoidance of service failures in the presence of faults as the basic mechanism for achieving dependability requirements [3]. We need to state here the difference between fault tolerance and high availability: a fault tolerant environment has no service interruption, while a highly available environment has minimal service interruption.

The key factor of fault tolerance (or fault transparency [4] is preventing failures due to system architectures and it addresses the fundamental characteristic of dependability requirements in two ways [5]:

- replication, i.e. providing multiple identical instances of the same component and choosing the correct result on the basis of a quorum (voting);
- redundancy, i.e. providing multiple identical instances of the same component and switching to one of the remaining instances in case of a failure (failover).

As a consequence, a system must be validated to ensure that its replication/redundancy mechanism has been correctly implemented and the system will provide the desired level of reliable service. Fault injection (the deliberate insertion of faults into a system to determine its response [6] [7]) offers an effective solution to this problem. Fault-injection experiments provide a means for understanding how these systems behave in the presence of faults (the monitoring of the effects the injected faults have on the systems final results). Simulated fault injection (or environment fault injection) [8] [9] [10] [11] can support all system abstraction levels – architectural, functional, logical, and electrical. This mixed-mode simulation, where the system is hierarchically decomposed for simulation at different abstraction levels, is particularly useful in the case of complex distributed systems.

———————————————

- *Andrey A. Shchurov - Department of Telecommunications Engineering, Czech Technical University in Prague, The Czech Republic, E-mail: shchuand@fel.cvut.cz*
- *Radek Mařík - Department of Telecommunications Engineering, Czech Technical University in Prague, The Czech Republic, E-mail: marikr@fel.cvut.cz*

In turn, strategies for the fault-injection experiments are generally based on methods for assessing system reliability (identifying potential faults and determining the resulting error effects) [12] [13] [14] [15]. Typically, it is a document-centric evaluation, where a group of engineers evaluates the system. But in the case of complex or non-standard systems, personal experience and/or intuition are often inadequate. Our main goal is the automated design and generation of testing procedures/specifications and plans for distributed systems based on end-user requirements and technical specifications as a necessary part of project documentation. Thus, to accomplish such a goal we need to identify a test strategy for fault-injection experiments based on a formal model with the following criteria: (1) it has to cover all aspects of distributed systems [4]; and (2) it has to be simple enough for commercial application.

This paper presents a systematic approach for identifying:
- characteristic sets of critical elements (hardware and software) of the distributed system under test (DSUT) for dependability testing (single points of failure and recovery groups) based on the concept of layered networks;
- the most important combinations of components from each recovery group based on the combinatorial (or truth tables) technique.

Based on these combinations, we determine the set of test templates which should be performed to demonstrate that protection mechanisms for achieving dependability requirements (fault tolerance or high availability) have been implemented correctly.

The rest of this paper is structured as follows. Section 2 introduces the related work. Section 3 presents a test strategy for fault-injection experiments based on the formal multilayer model of distributed systems for checklist generation missions and analytical tools for system reliability assessment. Section 4 introduces an example based on a simple multi-layered system. Finally, conclusion remarks are given in Section 5.

## 2 RELATED WORK

Nowadays, models for assessing reliability of distributed systems can be roughly classified into [16]:
- user-centric models;
- architecture-based models;
- state-based models.

### 2.1 User-Centric Models

Generally, user-centric models can be defined as the top-down or service-oriented approach (i.e. the viewpoint of the business/end-users community) to the reliability of distributed systems [17] [18] [19]. As reliability of any system has direct impact on the system usage, so these models focus on user/subscriber and provider behavior and basically work on the principle of evaluating transmission time to compute the execution time of each file or program under real conditions running in a distributed environment. As a consequence, the system reliability is based on the operational or usage profile of the given set of services.

The common analytical tool for user-centric models is time-based models (founded on the queueing theory) [19] [20]. User-centric approaches can be characterized as multi-stage problem solving processes where the system is conceived in terms of user behavior.

### 2.2 Architecture-Based Models

In contrast to the user-centric models, architecture-based models can be defined as the bottom-up or hardware-based approach (the viewpoint of the engineering community) to the reliability of distributed systems. In turn, they can be classified into: (1) component-oriented models; (2) communication-oriented models.

Component-oriented models represent distributed systems as a composition of multiple processors but completely ignore the failures of communications and assume that the communication channels (links) among the processors are perfect [21] [22] [23]. Without considering communication failures, the exchanged information between components (software and hardware) must always be correct. In this case, the problem of distributed system reliability can be reduced to a parallel-series structure. In turn, the parallel-series reliability is easy to calculate [23] [24] [25] [26]. Such condition may be a good approximation for a system that exchanges only a little information among nodes, such as those where the processors do only their own jobs (no intensive data transmission).

The analytical tool for component-oriented models is reliability block diagrams (one of the conventional and most common tools of system reliability analysis [25] [26]).

In contrast to the component-oriented models, *communication-oriented models* consider the communication failures and assume that the components themselves (the nodes of networks) are always perfect [23] [26] [27]. They suppose that the system failures are caused by the communication failures on channels (links) while the components (or nodes) cannot fail during the executing of programs. Such condition is a good approximation for cases where the communication time dominates the time of program execution or the components are highly reliable in comparison to the channels.

The analytical tool for communication-oriented models is network diagrams (commonly used in representing communication networks consisting of individual links [23]).

An additional effective analytical tool for architecture-base models is fault tree diagrams (the underlying graphical model in fault tree analysis) [23] [25] [28]. Whereas the reliability block diagrams and network diagrams are mission success oriented, the fault tree shows which combinations of the component failures can lead to system failures. And fault tree diagrams can describe the fault propagation in systems. However, repair and maintenance (two important operations in system analysis) cannot be expressed using a fault tree formulation.

1167

## 2.3 State-Based Models

The first generation of state-based models that considered both node failures and link failures have a common assumption - the operational probabilities of nodes or links are constant without considering bandwidth and content (constant-reliability models) [29] [30] [31]. However, this assumption of the constant-reliability of elements is not suitable in practice. Intuitively, downloading a larger file from a remote site will have a higher risk of failure than downloading a smaller file through the same link [32].

The most recent models relax this assumption for the elements (nodes and links). Instead, they assume that the failures of elements follow Poisson processes, so that the more time an element works (including execution and communication), the less reliable that element is [32] [33] [34]. In addition, the traditional models study the network topology by physical links and nodes that are static without considering dynamic changes of components and logic structures. To solve these problems, recent models use a virtual structure instead of physical structure [33] [34].

The analytical tool for state-based models is Markov models [23] [25]. To deal with all sorts of errors such as time-out failures, blocking failures, network failures, etc. (which can occur during operations of execution and communication), a hierarchical model must be used. This model suggests tackling various errors in different layers and uses Markov state principle to map layers into different physical states [16].

In turn, the general approach (common to all types of models) is to treat reliability as a complex problem and to decompose the distributed system into a hierarchy of related subsystems or components. Rebaiaia and Ait-Kadi [35] provide a survey of methods, algorithms and software tools. But it is important to note that the reliability evaluation problem is NP-complete and, as a consequence, the generation of an exact solution is very problematic. An interesting solution called mission-oriented reliability is represented by Wang et al [36] and Luo et al [37] based on the concept of layered networks [38]. The core component of this solution is a two-layer model with the lower-layer topology (physical network) for a physical graph and the upper-layer topology (mission network) for a logical graph. Both graphs have the same nodes and one edge in the logical graph is related to a path from the source node to the destination node in the physical graph. In that way, the network missions are modeled as a network, and the mission-oriented network reliability can be calculated on the mission network which consists of many fewer nodes and links than the physical network does (reduction of complexity).

Moreover, reliability of network topologies is the great challenge for all these models. Network communications are usually considered either (1) as physical communication structures based on the properties of communication hardware (physical links and nodes); or (2) as virtual communication structures (virtual information links) which normally hide the properties of communication hardware. In both cases, the layered structure of real communication protocols is completely ignored.

## 3 TEST STRATEGY FOR FAULT-INJECTION EXPERIMENTS

The essential idea of our approach is based on the concept of layered complex networks [38]. But in contrast to [36] [37], the core component of our solution is the formal four layered model for test generation missions [39]. This model is stated as a four-layered graph as follows:

- The functional (or ready-for-use system) architecture layer defines functional components and their interconnections (end-user requirements representation). An important note – this layer can be used for representation of social networks.
- The service architecture layer defines software-based components (services/applications) and their interconnections.
- The logical architecture layer defines logical (virtual) components and their interconnections.
- The physical architecture layer defines hardware (physical) components and their interconnections.
- The interlayer projections define all types of components hierarchical (interlayer) relations/mapping. These relations make the layered model consistent and represent interlayer technologies (virtualization, clustering, etc.) used to build DSUT.

The model formal notation $G_n$ for each layer n can be represented as:

$$G_n = (V_n, E_n, M_{n-1}^n, V_{n-1}) \qquad (1)$$

And:

$$G = \bigcup_{n=1}^{N} G_n \qquad (2)$$

where G is multi-layered 3D graph, derived from the system specification; N is the number of DSUT layers; $V_n$ is a finite, non-empty set of components on layer *n*; $E_n$ is a finite, non-empty set of component-to-component connections on layer *n*; $M_{n-1}^n$ is a finite set of component-to-component projection from layer *n* to layer *n-1*; and $V_{n-1}$ is a finite set of components on layer *n-1*.

Applying the requirements-coverage test strategy [40] to the model covers each interaction from the end-user requirements on system, logical and physical architectural layers and, as a consequence, provides the sets of test templates for each architectural layer:

$$T_n = \{(P_{n1}, c_{n1}), (P_{n2}, c_{n2}), \dots, (P_{nk}, c_{nk})\} \qquad (3)$$

where $P_{ni}$ represents the paths (or data flows) in $G_n$; and $c_{ni}$ are technical characteristics of component-to-component communication processes. In turn, each path $P_{ni}$ is the set of individual components which communicate each other and define this path (data flow):

$$P_{ni} = \{v_{n1}, v_{n2}, \dots, v_{nm}\}, \ v_{ni} \in V_n \qquad (4)$$

This approach allows use of the advantages of (1) the concept of layered complex networks [38] and (2) the approach of mission-oriented reliability [37] but it covers all layers of OSI Reference Model [41] and, as a consequence, both software-based and network-based aspects of distributed systems [39].

The next steps are based on the appropriate analytical tools for reliability assessment (analysis is performed independently for each architectural layer):

- success (logic) tree approach (as a special case of the fault tree approach);
- combinatorial (or truth tables) technique for the logic trees evaluation.

### 3.1 Success Tree Approach

The system performance can be considered from two opposite viewpoints: the various ways that a system fails or the various ways a system succeeds. The success tree approach [28] is a deductive process by means of which a desirable event, called the top event, is postulated, and the possible ways for this event to occur are systematically deduced. The success tree, which shows the various combinations success events that guarantee the occurrence of the top event, can be logically represented by path sets. The Boolean expression for the success tree can be written as [42]:

$$S_n = P_{n1}^* \vee P_{n2}^* \vee \ldots \vee P_{nk}^* \qquad (5)$$

where $S_n$ is the top event which denotes the state of the system on layer n (the entire system is in operational state iff it is in operational state on all layers simultaneously):

$$S_n = \begin{cases} 1, & \text{system is in operational state (OS)} \\ 0, & \text{system is in failure state (FS)} \end{cases} \qquad (6)$$

and $P_{ni}^*$ represents the path sets of the logical tree on layer $n$. In turn, each path set can be written as [42]:

$$P_{ni}^* = v_{n1}^* \wedge v_{n2}^* \wedge \ldots \wedge v_{nm}^* \qquad (7)$$

where $v_{ni}^*$ represents the basic events (or the state of individual components) in the success tree on layer n:

$$v_{ni}^* = \begin{cases} 1, & \text{component } v_{ni} \text{ is in OS} \\ 0, & \text{component } v_{ni} \text{ is in FS} \end{cases}, v_{ni} \in V_n \qquad (8)$$

In the case of complex systems that need more than one desirable top event simultaneously (composed of subsystems), the resulting expression of the entire system can be defined as a conjunction of success trees of their subsystems.

One of the main purposes of representing logical trees in terms of Boolean equations is that these equations can be reduced to its the most compact form which represents the minimal path sets or "minimal prevention sets" [42]. In our case, the most convenient representation of these compact forms is the conjunctive normal form (CNF) that provides a particular representation of the success tree as a set of sets of basic events. The Boolean expression of the success tree in conjunctive normal form for each layer n can be written as:

$$S_n^{CNF} = \wedge_{i=1}^{l} \left( \vee_{j=1}^{r_i} v_{nij}^* \right) \qquad (9)$$

where $l$ is the number of clauses in the CNF expression, and $r_i$ is the number of literals (or basic events) in each clause; and $v_{nij}^*$ represents the basic events:

$$v_{nij}^* = \begin{cases} 1, & \text{component } v_{nij} \text{ is in OS} \\ 0, & \text{component } v_{nij} \text{ is in FS} \end{cases}, v_{nij} \in V_n \qquad (10)$$

And:

$$\{v_{nij} \mid 1 \leq i \leq l; 1 \leq j \leq r_i\} \subseteq \{v_{ni} \mid 1 \leq i \leq m\} \subseteq V_n \qquad (11)$$

This resulting CNF expression defines two characteristic (prevention) sets for dependability testing:

*Single Points of Failure* [13]. If a clause of the resulting expression in CNF is a literal, i.e. it represents a single component of DSUT, so this component is a single point of failure. It means all possible paths can be destroyed by removing this component. As a consequence, components of that kind do not need additional testing but disaster recovery plans as part of project documentation. The set of components that are single points of failure for each layer *n*:

$$SPOF_n = \{v_{nij} \mid 1 \leq i \leq l; 1 \leq j \leq r_i; r_i = 1\}, v_{nij} \in V_n \qquad (12)$$

*Recovery Groups* [13]. If a clause of the resulting expression in CNF is a disjunction of literals, i.e. it represents a group of component of DSUT, so this group provides topological redundancy. It means all possible paths cannot be destroyed by removing a single component of this group – an alternative path (or paths) still exists. As a consequence, components of that kind need additional testing of protection mechanisms (sensing and switching). The set of components that provide fault tolerance for each layer *n*:

$$RG_n = \{v_{nij} \mid 1 \leq i \leq l; 1 \leq j \leq r_i; r_i \geq 2\}, v_{nij} \in V_n \qquad (13)$$

For the case of systems which can tolerate failures of *k* arbitrary components simultaneously:

$$l \geq k, \ r_i \geq k+1 \qquad (14)$$

A special case of recovery groups is the set of access points or end-user components (hardware and software). As a rule, these components are starting points of the most paths (data flows). However, generally fault tolerant design (replication or redundancy) is normally not used for end-user components. This statement is based on two main reasons:

- economic reason – replication/redundancy tends to increase the system cost;
- technical reason – replication/redundancy may increase complexity to the point where the replication/redundancy itself contributes to accidents [1].

As a consequence, DSUT components of that kind do not contain protection mechanisms and do not need additional testing. Based on this assumption, access points (end-user components) can be eliminated from analysis by converting all literals which represent these components into tautologies. In this case, the resulting Boolean structures for each layer n can be represented as:

$$S_n^{CNF} = \wedge_{i=1}^{l}\left(\vee_{j=1}^{r_i} v_{nij}^{**}\right), \quad v_{nij}^{**} = \begin{cases} 1, & v_{nij} \in A_n \\ v_{nij}^{*}, & v_{nij} \notin A_n \end{cases}, A_n \subset V_n \quad (15)$$

where $A_n$ is the set of access points (or end-user devices) on layer n.

### 3.2 Logic Trees Evaluation

When calculating the probability of multiple simultaneous failures, the number of combinations that should be tested can be reduced dramatically. So, for further analysis of the resulting Boolean structures in CNF we can use the combinatorial (or truth tables) technique [25]. This method relies on a combinatorial algorithm to exhaustively generate all probabilistically significant combinations of both failure and success events and subsequently to propagate the effect of each combination on the logic tree to determine the state of the top event. The quantification of logic trees based on the combinatorial method yields exact results and these are associated with a specific physical state of DSUT [25].

The sum of the probabilities of all possible combinations is unity because the combinations are all mutually exclusive and cover all event space [25]. For the case of independent identical units (IIU) with reliability of $p$, the universal set for DSUT can be represented as:

$$\sum_{k=0}^{m}\binom{m}{k}(1-p)^k p^{m-k} = 1 \quad (16)$$

where k is the number of failed components, and m is the number of DSUT individual components defined by the resulting Boolean structure in CNF – see (15):

$$m = \sum_{i=1}^{l} r_i \quad (17)$$

Theoretically testing activities should cover the universal set. However, in this case the number of combinations C that should be tested is:

$$C = \sum_{k=0}^{m}\binom{m}{k} = 2^m \quad (18)$$

For independent identical units (IIU) with reliability of $p$, the reliability expression $R$ for DSUT can be defined based on the set of components that provide fault tolerance (or clauses of the resulting CNF expression – see (15)):

$$R = \prod_{i=1}^{l}\left(\sum_{j=0}^{r_i-k_i}\binom{r_i}{j}(1-p)^j p^{r_i-j}\right) \quad (19)$$

where $k_i$ is the number of individual components in each recovery group of the $r_i$ items that must be in the operational state for the system to be in the operational state (*k-out-of-r* structure).

In the worst case of $k_i = 1$ (this case determines the maximum number of possible combinations), the number of combinations in each recovery group is:

$$C_i = 2^{r_i} - 1 \mid r_i \geq 2 \quad (20)$$

However one combination must be excluded from analysis – all components of a system are in operational state – this state is the initial state for fault-injection testing. Thus, the number of combinations $C$ that must be tested is:

$$C = \left[\prod_{i=1}^{l}(2^{r_i} - 1)\right] - 1 \quad (21)$$

Obviously, this result is better than the universal set coverage. Nevertheless for a large volume of $l$ and/or $r_i$, the generation of this large number of tests is still impractical and, as a consequence, some additional steps (optimization) must be used.

The next step of our approach is based on two basic assumptions:

- Real distributed engineering systems are usually under great financial and timing constraints and, as a consequence, they consist of the smallest possible set of components with a minimal number of communication links as a tradeoff between cost and reliability requirements, i.e. their topologies can be represented by Harary graphs [43] (connected simple graphs with a minimal number of edges) with the additional links based on technological requirements.
- It is not necessary to cover all possible successful combinations but the most important (the most likely) only [44]. In turn, the most important combination can be derived from end-user requirements as the number of failures which a system is able to tolerate simultaneously.

We need to state here:

- Big data centers which can contain thousands of servers (hardware and software instances) are beyond the scope of this work due to possible legislation challenges [45].

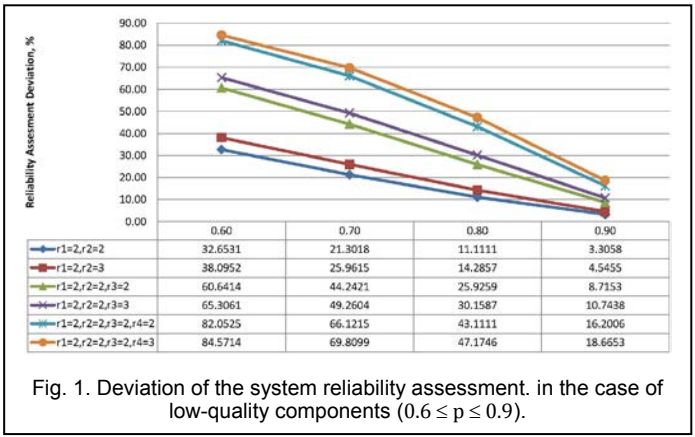

Fig. 1. Deviation of the system reliability assessment. in the case of low-quality components ($0.6 \leq p \leq 0.9$).

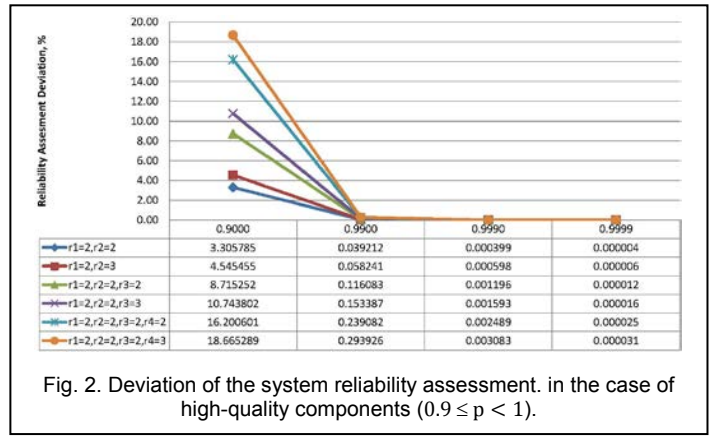

Fig. 2. Deviation of the system reliability assessment. in the case of high-quality components ($0.9 \leq p < 1$).

Nevertheless, they can be divided into small subsystems based on individual functional tasks (end-user requirements).

- Specific areas (the military, nuclear or aerospace industries) are beyond the scope of this work.

The most typical values of the number of clauses (after elimination of end-user components) are based on architectural solutions:

- The physical architecture layer. The assumption is based on the hierarchical design model [46]. The upper bound represents core, distribution (aggregation) and access network layers with an additional layer for server hardware and the lower bound represents the fact that even the simplest architecture consist of at least one network component and at least one server: $2 \leq l \leq 4$.
- The logical architecture layer. It is similar to the physical architecture layer but based on virtual components: $2 \leq l \leq 4$.
- The service architecture layer. The assumption is based on the three-tiered architecture [4]. The upper bound represents user-interface, processing and data levels and the lower bound represents the fact that in the case of the simplest client-server model clients can directly communicate with servers: $1 \leq l \leq 3$.

In turn, the most typical values of the number of literals (individual components) in each clause (after elimination of end-user components) are based on technological solutions:

- Redundancy [5]. Two identical instances of the same component in active/standby (switching to the remaining instances in case of a failure) or active/active mode, i.e. 1-out-of-2 structure: $r_i = 2$, $k_i = 1$.
- Replication [5]. Three identical instances of the same component in active mode (choosing the correct result on the basis of a quorum), i.e. 2-out-of-3 structure: $r_i = 3$, $k_i = 2$.

The typical reliability values (examples) of COTS equipment (hardware) is shown in Table 1. Unfortunately, it is impossible to find similar data for COTS software. Nevertheless, for stable releases of software components (services and/or platforms – operating systems/firmware) the reliability function becomes dominated by hardware failures and the impact of software failures becomes smaller with respect to the total component failure rate [26] [47]. So, analysis of the physical architecture layer can be applicable to the entire system.

1. In the case of a system which can tolerate failures of a single arbitrary component (or just have no single points of failures), the most important states that should be covered by tests are:
   - a system is in operational state and all components are in operational state;
   - a system is in operational state and a single arbitrary component is in failure state.

A special case (or an exception) is if solutions are based on virtualization technologies: they must tolerate failures of a single arbitrary physical server (hardware failure) while another one is in maintenance mode [48]. However it is a question of the resources sharing/allocation, not specific protection mechanisms.

The system reliability assessment $R_{L1}$ (the probability of finding DSUT in these states) is:

$$R_{L1} = p^m + \sum_{i=1}^{l}[r_i(1-p)p^{r_i-1}]p^{m-r_i} = p^m + m(1-p)p^{m-1} \quad (19)$$

In turn, the deviation of the system reliability assessment D can be defined as:

$$D = \frac{R - R_{L1}}{R} 100\% \quad (20)$$

In the case of low-quality individual components ($0.6 \leq p \leq 0.9$), the result is shown in Fig. 1. In turn, Fig. 2 shows the result of high-quality components ($0.9 \leq p < 1$). Based on information from Table 1, the average value of the system reliability assessment deviation in the case of COTS equipment is less than 1%.

As a consequence, the number of combinations C that should be tested based on the system reliability assessment is quite trivial compared with the universal set coverage:

TABLE 1
RELIABILITY OF COTS EQUIPMENT (EXAMPLES)

| N/N | Components | MTTF, hours * | Component Reliability | | |
|---|---|---|---|---|---|
| | | | 8760 hours | 17520 hours | 26280 hours |
| 1 | Cisco Catalyst C3650-48P | 383 760 | 0.9774 | 0.9554 | 0.9338 |
| 2 | Cisco Catalyst C3650-24T | 661 800 | 0.9869 | 0.9739 | 0.9611 |
| 3 | Cisco Catalyst 2960X-48FPD-L | 233 370 | 0.9632 | 0.9277 | 0.8935 |
| 4 | Cisco Catalyst 2960X-24TS-LL | 622 350 | 0.9860 | 0.9722 | 0.9587 |
| 5 | Cisco Aironet 1140 Access Point | 390 000 | 0.9778 | 0.9561 | 0.9348 |
| 6 | Dell PowerEdge 6450 Server | 45 753 | 0.8258 | 0.6819 | 0.5630 |
| 7 | Plextor PX-512M5P SSD | 2 400 000 | 0.9964 | 0.9927 | 0.9891 |

* *Data sources: Cisco - [49] [50] [51]; Dell - [52]; Plextor - [53].*

$$C = m, \ m \geq 2 \quad (21)$$

But this trivial result covers at least 99% of specific states of DSUT.

2. In the specific case of reliable systems which can tolerate failures of up to two arbitrary components simultaneously, the most important states that should be covered by tests are:
- a system in operational state and all components of a system are in operational state;
- a system in operational state and a single arbitrary component is in failure state;
- a system in operational state and two arbitrary components are in failure state.

So, the system reliability assessment $R_{L2}$ can be represented as:

$$R_{L2} = p^m + m[(1-p)p^{m-1}] + \frac{m(m-1)}{2}[(1-p)^2 p^{m-2}] \quad (22)$$

In this case, the number of combinations that should be tested C is:

$$C = m + \frac{m(m-1)}{2} = \frac{m(m+1)}{2}, \ m \geq 3 \quad (23)$$

It might be difficult to say whether this result is acceptable for commercial application. A possible solution is using two simple (which can tolerate failures of a single arbitrary component) systems in parallel instead of a complex one.

### 3.3 Fault-Injection Experiments

Based on its nature, the dependability testing (or testing of the sensing and switching protection mechanisms) must include two main steps:

*Component Fault Injection (FIJ).* This step determines two interrelated activities that cannot be divided:
- sensing mechanism verification – the sensing mechanism is able to (1) detect a component failure, and (2) trigger the switching mechanism;

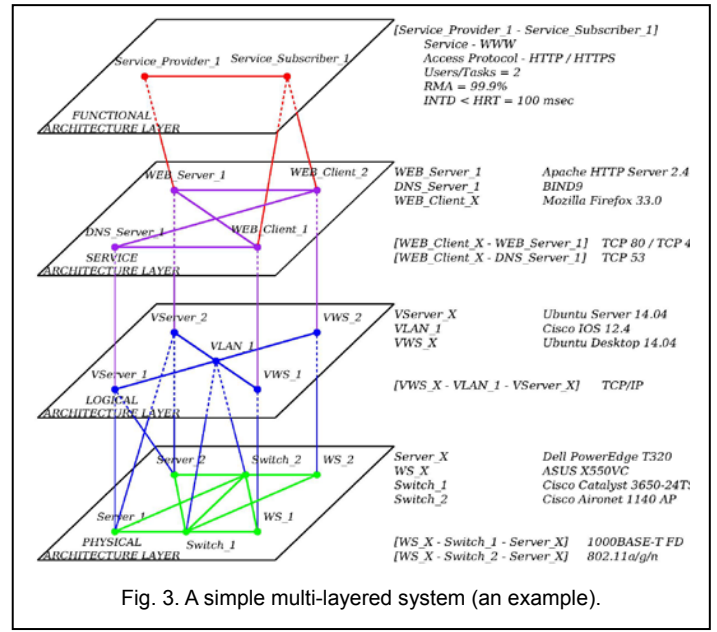

Fig. 3. A simple multi-layered system (an example).

- switching mechanism verification – the switching mechanism is able to reconfigure DSUT topology (re-route data flows) due to the component failure.

*Component Repair (FIJ).* In turn, this step determines the two interrelated activities:
- sensing mechanism verification – the sensing mechanism is able to (1) detect a component *resurrection*, and trigger the switching mechanism (if necessary);
- switching mechanism verification – the switching mechanism is able to restore DSUT initial topology (if necessary).

As a consequence, each fault injection action/step defines two test templates and the set of test templates $T_n^{FIJ}$ for fault injection experiments can be defined as:

$$T_n^{FIJ} = \{(FIJ(v_{nij}), \overline{FIJ}(v_{nij})) \mid 1 \leq i \leq l; \ 1 \leq j \leq r_i; r_i \geq 2\} \quad (24)$$

So, in the case of systems which can tolerate failures of a single arbitrary component, the number of test templates for fault injection experiments (dependability testing) $|T^{FIJ}|$ is:

$$|T^{FIJ}| = \sum_{n=1}^{N} 2(|RG_n| - |A_n|) = \sum_{n=1}^{N} 2(|V_n| - |SPOF_n| - |A_n|) \quad (25)$$

The next step is based on the following assumptions:
- The number of DSUT layers is limited by the system model (see [39]): $N = 3$ – the Functional architecture layer does not define real (software or hardware) components.

TABLE 2.1
APPLICATION OF THE REQUIREMENTS-COVERAGE STRATEGY.
BASIC SUBSYSTEM (END-USER REQUIREMENT):
[SERVICE_SUBSCRIBER_X, SERVICE_PROVIDER_X]

| N/N | Model layer | Data Flows |
|---|---|---|
| 1 | Functional | [Service_Subscriber_1, Service_Provider_1] |
| 2 | Service | [WEB_Client_1, WEB_Server_1]<br>[WEB_Client_2, WEB_Server_1] |
| 3 | Logical | [VWS_1, VLAN_1, VServer_2]<br>[VWS_2, VLAN_1, VServer_2] |
| 4 | Physical | [WS_1, Switch_1, Server_1]<br>[WS_1, Switch_1, Switch_2, Server_1]<br>[WS_1, Switch_2, Server_1]<br>[WS_1, Switch_2, Switch_1, Server_1]<br>[WS_1, Switch_1, Server_2]<br>[WS_1, Switch_1, Switch_2, Server_2]<br>[WS_1, Switch_2, Server_2]<br>[WS_1, Switch_2, Switch_1, Server_2]<br>[WS_2, Switch_1, Server_1]<br>[WS_2, Switch_1, Switch_2, Server_1]<br>[WS_2, Switch_2, Server_1]<br>[WS_2, Switch_2, Switch_1, Server_1]<br>[WS_2, Switch_1, Server_2]<br>[WS_2, Switch_1, Switch_2, Server_2]<br>[WS_2, Switch_2, Server_2]<br>[WS_2, Switch_2, Switch_1, Server_2] |

TABLE 2.2
APPLICATION OF THE REQUIREMENTS-COVERAGE STRATEGY.
ADDITIONAL SUBSYSTEM (TECHNICAL REQUIREMENT DERIVED FROM END-USER REQUIREMENT): [WEB_CLIENT_X, DNS_SERVER_X]

| N/N | Model layer | Data Flows |
|---|---|---|
| 1 | Functional | - |
| 2 | Service | [WEB_Client_1, DNS_Server_1]<br>[WEB_Client_2, DNS_Server_1] |
| 3 | Logical | [VWS_1, VLAN_1, VServer_1]<br>[VWS_2, VLAN_1, Vserver_1] |
| 4 | Physical | [WS_1, Switch_1, Server_1]<br>[WS_1, Switch_1, Switch_2, Server_1]<br>[WS_1, Switch_2, Server_1]<br>[WS_1, Switch_2, Switch_1, Server_1]<br>[WS_1, Switch_1, Server_2]<br>[WS_1, Switch_1, Switch_2, Server_2]<br>[WS_1, Switch_2, Server_2]<br>[WS_1, Switch_2, Switch_1, Server_2]<br>[WS_2, Switch_1, Server_1]<br>[WS_2, Switch_1, Switch_2, Server_1]<br>[WS_2, Switch_2, Server_1]<br>[WS_2, Switch_2, Switch_1, Server_1]<br>[WS_2, Switch_1, Server_2]<br>[WS_2, Switch_1, Switch_2, Server_2]<br>[WS_2, Switch_2, Server_2]<br>[WS_2, Switch_2, Switch_1, Server_2] |

- The maximum possible number of tests templates (dependability testing) can be achieved if: (1) the number of end-users (access point) which must be eliminated from analysis has the minimal value: $|A_n| = 1; 1 \leq n \leq 3$; and (2) the number of single points of failure has the minimal value: $|SPOF_n| = 0; 1 \leq n \leq 3$.

So:

$$|T^{FIJ}| \leq \sum_{n=1}^{3} 2(|V_n| - 1) < \sum_{n=1}^{3} 2|V_n| \qquad (26)$$

In turn, in the specific case of systems which can tolerate failures of up to two arbitrary components simultaneously, the number of tests templates for fault injection experiments is:

$$|T^{FIJ}| \leq \sum_{n=1}^{3} 2(|V_n| - 1)^2 < \sum_{n=1}^{3} 2|V_n|^2 \qquad (27)$$

## 4 CASE STUDY

As a practical example, we have a simple multi-layered system (see Fig. 3). The requirements-coverage strategy application to this simple example defines data flows which cover the model based on requirements – see Table 2.1 and Table 2.2.
The resulting Boolean structures in conjunctive normal form after elimination of end-user components for each layer are shown in Table 3. In turn, these structures define characteristic sets for dependability testing:
- two sets of single points of failure:
Service layer:
$$SPOF_3 = \{WEB\_Server\_1, DNS\_Server\_1\}$$
Logical layer:

$$SPOF_2 = \{VLAN\_1, VServer\_1, VServer\_2\}$$

- the set of components that provide fault tolerance (a recovery group):
Physical layer:
$$RG_1 = \{Switch\_1, Switch\_2, Server\_1, Server\_2\}$$

As a result we have a set of objects which need additional (dependability) testing of sensing and switching protection mechanisms.

A formal set of test templates covers all successful DSUT operation status - see Table 4 NN 1 – 3, 5 – 7 and 9 – 11. The system reliability is 96.93723651%.

In turn, the optimized set of test templates covers the most important only - see Table 4} NN 1 – 3, 5 and 9. The system reliability assessment is 95.93835902%.

So, based on these optimization steps, we can reduce the number of tests templates from 20 to 8 (2.5 times) with the Reliability Assessment Deviation 1.030%.

## 4 CONCLUSION

Deployment of distributed systems sets high requirements for procedures, tools and approaches for complex testing of these systems. And the most difficult part of systems deployment is the question of assurance that system dependability protection mechanisms (fault tolerance or high availability) have been implemented correctly and a system is able to provide the desired level of reliable service.

This paper presents a systematic approach for identifying critical elements based on the concept of layered networks [38]. The key component is the formal four layered model for test generation missions [39]. This model is the four-layered 3D graph, derived from the system technical specifications, which covers all layers of OSI Reference Model [41].

TABLE 3
RESULTING BOOLEAN EXPRESSIONS FOR DEPENDABILITY ANALYSIS

| N/N | Model layer | Boolean Expression $A \wedge B \wedge (C \vee D)$ | Characteristic Sets — Single Points of Failures $A, B$ | Characteristic Sets — Recovery Groups $C, D$ |
|---|---|---|---|---|
| 1 | Functional | - | - | - |
| 2 | Service | WEB_Server_1 ∧ DNS_Server_1 | WEB_Server_1, DNS_Server_1 | - |
| 3 | Logical | VLAN_1 ∧ VServer_1 ∧ VServer_2 | VLAN_1, VServer_1, VServer_2 | - |
| 4 | Physical | (Switch_1 ∨ Switch_2) ∧ (Server_1 ∨ Server_2) | - | Switch_1, Switch_2, Server_1, Server_2 |

Applying the requirements-coverage test strategy [40] to the model covers each interaction from the end-user requirements on system, logical and physical architectural layers and, as a consequence, provides the sets of paths (or data flows) for each layer (each path is the set of individual components which communicate each other and define this path).

The next steps are based on the analytical tools for reliability assessment (analysis is performed independently for each architectural layer):

- Success (logic) tree approach (as a special case of the fault tree diagrams) allows us to represent the sets of paths (or data flows) as a Boolean structure in conjunctive normal form (CNF) and, as a consequence, defines two characteristic sets for dependability testing: (1) sets of single points of failure; (2) sets of components that provide fault tolerance (recovery groups).
- In turn, combinatorial (or truth tables) technique for the logic trees evaluation defines the most important combinations of components from each recovery group that must be tested.

Based on these combinations, we determine the set of test templates which should be performed to demonstrate that protection mechanisms for achieving dependability requirements (fault tolerance or high availability) have been implemented correctly.

This approach allows use of the advantages of (1) the concept of layered complex networks and (2) the approach of mission-oriented reliability – reduction of complexity - but it covers all layers of OSI Reference Model and, as a consequence, both software-based and network-based aspects of distributed systems.

TABLE 4
COMBINATORIAL METHOD (TRUTH TABLE)

| N/N | Server_1 | Server_2 | Switch_1 | Switch_2 | System Operation Status | Probability of Operation Status * |
|---|---|---|---|---|---|---|
| 1 | 1 | 1 | 1 | 1 | OS | 0.6580712823 |
| 2 | 1 | 1 | 1 | 0 | OS | 0.0149408698 |
| 3 | 1 | 1 | 0 | 1 | OS | 0.0087351645 |
| 4 | 1 | 1 | 0 | 0 | FS | 0.0001983234 |
| 5 | 1 | 0 | 1 | 1 | OS | 0.1388181368 |
| 6 | 1 | 0 | 1 | 0 | OS | 0.0031517311 |
| 7 | 1 | 0 | 0 | 1 | OS | 0.0018426564 |
| 8 | 1 | 0 | 0 | 0 | FS | 0.0000418357 |
| 9 | 0 | 1 | 1 | 1 | OS | 0.1388181368 |
| 10 | 0 | 1 | 1 | 0 | OS | 0.0031517311 |
| 11 | 0 | 1 | 0 | 1 | OS | 0.0018426564 |
| 12 | 0 | 1 | 0 | 0 | FS | 0.0000418357 |
| 13 | 0 | 0 | 1 | 1 | FS | 0.0292832640 |
| 14 | 0 | 0 | 1 | 0 | FS | 0.0006648481 |
| 15 | 0 | 0 | 0 | 1 | FS | 0.0003887028 |
| 16 | 0 | 0 | 0 | 0 | FS | 0.0000088251 |
|   |   |   |   |   |   | **1.0000000000** |

\* *Probability of system operatin status is calculated based on data from Table 1.*

## REFERENCES

[1] N.G. Leveson, Safeware: system safety and computers, ACM, 1995.

[2] A. Avizienis, J.-C. Laprie, B. Randell and C. Landwehr, "Basic concepts and taxonomy of dependable and secure computing," *Dependable and Secure Computing, IEEE Transactions on,* vol. 1, no. 1, pp. 11-33, January 2004.

[3] A. Bucchiarone, H. Muccini and P. Pelliccione, "Architecting Fault-tolerant Component-based Systems: From Requirements to Testing," *Electron. Notes Theor. Comput. Sci.,* vol. 168, pp. 77-90, February 2007.

[4] A.S. Tanenbaum and M.v. Steen, Distributed Systems: Principles and Paradigms, 3rd ed., Prentice Hall Press, 2013.

[5] H. Langmaack, W.-P. d. Roever and J. Vytopil, Eds., Formal Techniques in Real-Time and Fault-Tolerant Systems: Third International Symposium Organized Jointly with the Working Group Provably Correct Systems, Springer-Verlag, 1994.

[6] J.A. Clark and D.K. Pradhan, "Fault injection: a method for validating computer-system dependability," *Computer,* vol. 28, no. 6, pp. 47-56, June 1995.

[7] D. Avresky, J. Arlat, J.-C. Laprie and Y. Crouzet, "Fault injection for formal testing of fault tolerance," *Reliability, IEEE Transactions on,* vol. 45, no. 3, pp. 443-455, September 1996.

[8] D.d. Andres, J.-C. Ruiz, D. Gil and P. Gil, "Fault Emulation for Dependability Evaluation of VLSI Systems," *Very Large Scale Integration (VLSI) Systems, IEEE Transactions on,* vol. 16, no. 4, pp. 422-431, April 2008.

[9] C. Bolchini, A. Miele and D. Sciuto, "Fault Models and Injection Strategies in SystemC Specifications," in *Digital System Design Architectures, Methods and Tools, 2008. DSD '08. 11th EUROMICRO Conference on*, 2008.

[10] D. Lee and J. Na, "A Novel Simulation Fault Injection Method for Dependability Analysis," *Design Test of Computers, IEEE,* vol. 26, no. 6, pp. 50-61, 2009.

[11] L. Entrena, M. Garcia-Valderas, R. Fernandez-Cardenal, A. Lindoso, M. Portela and C. Lopez-Ongil, "Soft Error Sensitivity Evaluation of

Microprocessors by Multilevel Emulation-Based Fault Injection," *Computers, IEEE Transactions on,* vol. 61, no. 3, pp. 313-322, 2012.

[12] O.P. Yadav, N. Singh and P.S. Goel, "Reliability demonstration test planning: A three dimensional consideration," *Reliability Engineering & System Safety,* vol. 91, no. 8, pp. 882-893, 2006.

[13] E. Bauer and R. Adams, Reliability and Availability of Cloud Computing, 1st ed., Wiley-IEEE Press, 2012.

[14] K. Benz and T. Bohnert, "Dependability Modeling Framework: A Test Procedure for High Availability in Cloud Operating Systems," in *Vehicular Technology Conference (VTC Fall), 2013 IEEE 78th*, 2013.

[15] S. Reiter, M. Pressler, A. Viehl, O. Bringmann and W. Rosenstiel, "Reliability assessment of safety-relevant automotive systems in a model-based design flow," in *Design Automation Conference (ASP-DAC), 2013 18th Asia and South Pacific*, 2013.

[16] W. Ahmed and Y.W. Wu, "A Survey on Reliability in Distributed Systems," *J. Comput. Syst. Sci.,* vol. 79, no. 8, pp. 1243-1255, December 2013.

[17] E. Huedo, R.S. Montero and I.M. Llorente, "Evaluating the reliability of computational grids from the end users point of view," *Journal of Systems Architecture,* vol. 52, no. 12, pp. 727-736, 2006.

[18] V. Cortellessa and V. Grassi, "Reliability Modeling and Analysis of Service-Oriented Architectures," in *Test and Analysis of Web Services*, L. Baresi and E. d. Nitto, Eds., Springer Berlin Heidelberg, 2007, pp. 339-362.

[19] Z. Zheng and M.R. Lyu, "Collaborative Reliability Prediction of Service-oriented Systems," in *Proceedings of the 32Nd ACM/IEEE International Conference on Software Engineering*, 2010.

[20] D.J. Chen, M.C. Sheng and M.S. Horng, "Real-time distributed program reliability analysis," in *Parallel and Distributed Processing, 1993. Proceedings of the Fifth IEEE Symposium on*, 1993.

[21] J.-C. Laprie and K. Kanoun, "X-ware reliability and availability modeling," *Software Engineering, IEEE Transactions on,* vol. 18, no. 2, pp. 130-147, 1992.

[22] Y.S. Dai, M. Xie, K.L. Poh and S.H. Ng, "A model for correlated failures in N-version programming," *IIE Transactions,* vol. 36, no. 12, pp. 1183-1192, 2004.

[23] M. Xie, K.L. Poh and Y.S. Dai, Computing System Reliability: Models and Analysis, Springer, 2004.

[24] M.L. Shooman, Reliability of Computer Systems and Networks: Fault Tolerance, Analysis, and Design, John Wiley & Sons, 2002.

[25] M. Modarres, M. Kaminskiy and V. Krivtsov, Reliability Engineering And Risk Analysis: A Practical Guide, 2nd ed., CRC Press, 2010.

[26] M.L. Ayers, Telecommunications System Reliability Engineering, Theory, and Practice, 1st ed., Wiley-IEEE Press, 2012.

[27] D. J. Chen and T. H. Huang, "Reliability analysis of distributed systems based on a fast reliability algorithm," *Parallel and Distributed Systems, IEEE Transactions on,* vol. 3, no. 2, pp. 139-154, 1992.

[28] K.G. Stephens, A Fault Tree Approach to Analysis of Behavioral Systems: An Overview, Distributed by ERIC Clearinghouse, 1974.

[29] P. Kumar, S. Hariri and S.C. Raghavendra, "Distributed program reliability analysis," *Software Engineering, IEEE Transactions on,* Vols. SE-12, no. 1, pp. 42-50, 1986.

[30] Y.S. Dai, M. Xie and K.L. Poh, "Reliability Analysis of Grid Computing Systems," in *PRDC*, 2002.

[31] Y.S. Dai, M. Xie, K.L. Poh and G.Q. Liu, "A study of service reliability and availability for distributed systems," *Rel. Eng. & Sys. Safety*, vol. 79, no. 1, pp. 103-112, 2003.

[32] Y.S. Dai, M. Xie and K.L. Poh, "Reliability of grid service systems," *Computers & Industrial Engineering*, vol. 50, no. 1-2, pp. 130-147, 2006.

[33] Y.S. Dai, Y. Pan and X. Zou, "A Hierarchical Modeling and Analysis for Grid Service Reliability," *IEEE Trans. Computers,* vol. 56, no. 5, pp. 681-691, 2007.

[34] Y.S. Dai, B. Yang, J. Dongarra and G. Zhang, "Cloud Service Reliability: Modeling and Analysis," 2010. Available: http://www.techrepublic.com/resource-library/whitepapers/

[35] M.-L. Rebaiaia and D. Ait-Kadi, "Network Reliability Evaluation and Optimization: Methods, Algorithms and Software Tools," CIRRELT, 2013.

[36] X. Wang, N. Huang, W. Chen and R. Li, "A new method for evaluating the performance reliability of communications network," in *Information Networking and Automation (ICINA), 2010 International Conference on*, 2010.

[37] Q. Luo, M. Chen, X. Yin and H. Deng, "Testing mission-oriented network reliability via hierarchical mission network," in *Quality, Reliability, Risk, Maintenance, and Safety Engineering (ICQR2MSE), 2012 International Conference on*, 2012.

[38] M. Kurant and P. Thiran, "Layered Complex Networks," *Phys. Rev. Lett.,* vol. 96, no. 13, April 2006.

[39] A.A. Shchurov, "A Formal Model of Distributed Systems For Test Generation Missions," *International Journal of Computer Trends and Technology (IJCTT)*, vol. 15, no. 6, pp. 128-133, September 2014.

[40] A.A. Shchurov and R. Marik, "A Formal Approach to Distributed System Tests Design," *International Journal of Computer and Information Technology (IJCIT)*, vol. 3, no. 4, pp. 696-705, July 2014.

[41] ISO/IEC, *ITU-T Rec. X.200 - ISO/IEC 7498:1994 Information technology - Open Systems Interconnection - Basic Reference Model*, 1994.

[42] M. Stamatelatos, W. Vesely, J. Dugan, J. Fragola, J. Minarick and J. Railsback, Fault Tree Handbook with Aerospace Applications, NASA, 2002.

[43] M.v. Steen, Graph Theory and Complex Networks: An Introduction, 1st ed., Maarten van Steen, 2010.

[44] K. Dooley, Designing Large Scale LANs, O'Reilly Media, 2001.

[45] T. Erl, R. Puttini and Z. Mahmood, Cloud Computing: Concepts, Technology & Architecture, 1st ed., Prentice Hall Press, 2013.

[46] Cisco Systems, "Campus Design Summary," 2014.

[47] K.V. Vishwanath and N. Nagappan, "Characterizing Cloud Computing Hardware Reliability," in *Proceedings of the 1st ACM Symposium on Cloud Computing*, 2010.

[48] Cisco Systems, "Data Center Technology Design Guide," 2013.

[49] Cisco Systems, "Cisco Catalyst 3650 Series Switches Data Sheet".

[50] Cisco Systems, "Cisco Catalyst 2960-X Series Switches Data Sheet".

[51] Cisco Systems, "Cisco Aironet 1140 Series Access Point Data Sheet".

[52] Dell, "Dell Power Solutions Whitepaper".

[53] Plextor, "Plextor M5 Pro Solid-State Drive Whitepaper".